\input amstex
\documentstyle{amsppt}
\magnification=\magstep1 \pagewidth{16 true cm} \pageheight{23,5
true cm}
\def\proclaimheadfont@{\smc}
\def\demoheadfont@{\smc}
\NoRunningHeads \expandafter\redefine\csname logo\string @
\endcsname{}
 \TagsOnRight
 \UseAMSsymbols
 \topmatter
\title DISCRETE MODELS OF THE SELF-DUAL AND ANTI-SELF-DUAL EQUATIONS
\endtitle
\author Volodymyr Sushch \endauthor
\affil Lviv, Ukraine; Koszalin, Poland
\endaffil
\abstract   In the case of a gauge-invariant discrete model of
Yang-Mills theory difference self-dual and anti-self-dual
equations   are constructed.
\endabstract
\address {\eightpoint Technical University of Koszalin, Raclawicka Str.,
 15-17, 75-620 Koszalin, Poland}
\endaddress
\email {\eightpoint sushch\@lew.tu.koszalin.pl}
\endemail
\endtopmatter
\document
\baselineskip 10 pt

\noindent {\bf 1.~Introduction}
\bigskip
In $4$-dimensional non-abelian
gauge theory the self-dual and anti-self-dual connections are the
most important extrema of the Yang-Mills action. Consider a
trivial bundle $P=\Bbb R^4\times G$, where $G$ is some Lie group.
We define a connection as some $\frak g$-valued 1-form $A$, where
 $\frak  g$ is the Lie algebra   of the group $G$ [5]. Then the  connection 1-form $A$
can be written as  follows
$$A=\sum_{a,\mu}A^a_{\mu}(x)\lambda_adx^{\mu},\tag 1$$
  where   ${\lambda_a}$ is the basis of the Lie algebra $\frak g$.
   The  curvature 2-form $F$ of the connection $A$ is given by
     $$
     F=dA+A\wedge A.\tag2
     $$
     We specialize straightaway to the choice $G=SU(2)$, then
     $\frak g=su(2)$. We define the covariant exterior
     differentiation operator $d_A$ by
     $$
     d_A\Omega=d\Omega+A\wedge\Omega+(-1)^{r+1}\Omega\wedge
A,\tag3
  $$
  where $\Omega$ is an arbitrary $su(2)$-valued $r$-form.
  Compare (2) and (3) we obtain the Bianchi identity
  $$
     d_AF=0.\tag4
     $$
The Yang-Mills action $S$ can be conveniently expressed (see [5,
p.~256]) in terms of the 2-forms $F$ and $\ast F$ as
$$S=-\int\limits_{\Bbb R^4}tr(F\wedge\ast F),$$
 where  $\ast$  is the adjoint operator
 (Hodge star operator). The Euler-Lagrange equations for the
 extrema of $S$ are
 $$
   d_A\ast F=0.\tag5
     $$
     Equations (4), (5) are called
the Yang-Mills equations [4]. These equations are non-linear
coupled partial differential equations containing quadratic and
cubic terms in $A$.

In more traditional form the Yang-Mills equations are expressed in
terms of components of the connection $A$ and the curvature $F$
(see [2,3]). Let $$A_\mu=\sum_\alpha
A^\alpha_{\mu}(x)\lambda_\alpha$$ be the component of the
connection 1-form (1). Then the components of the curvature form
are given by $$F_{\mu\nu}=\frac{\partial A_{\nu}}{\partial
x^\mu}+\frac{\partial A_{\mu}}{\partial x^\nu}+[A_\mu, A_\nu],$$
where $[\cdot \ ,\cdot]$ be the commutator of the algebra Lie
$su(2)$. In local coordinates the covariant derivative $\nabla_j$
can be written $$\nabla_j F_{\mu\nu}=\frac{\partial
F_{\mu\nu}}{\partial x^j}+[A_j, F_{\mu\nu}].$$ Then we can write
Equations (4), (5) as $$ \nabla_j F_{\mu\nu}+\nabla_\mu F_{\nu
j}+\nabla_\nu F_{j\mu}=0,\tag6$$ $$\sum_{\mu=1}\frac{\partial
F_{\mu\nu}}{\partial x^\mu}+[A_\mu, F_{\mu\nu}]=0.\tag7$$ Note
that Equations 7 are obtained in the case of Euclidean space $\Bbb
R^4$.

 The self-dual and
anti-self-dual connections are solutions of the following
 nonlinear first order differential equations
 $$
 F=\ast F, \qquad F=-\ast F.\tag 8
 $$
 Equations (8) are called self-dual and anti-self-dual
 respectively. It is obviously that if one can find $A$ such that
 $F=\pm\ast F$, then the Yang-Mills equations (5) are automatically
 satisfied.
 \bigskip
\bigskip
\noindent {\bf 2.~The discrete model in $\Bbb R^4$}
\bigskip
 In [6] the gauge invariant discrete model of the Yang-Mills
 equations is constructed in the case of the $n$-dimensional Euclidean space
 $\Bbb R^n$. Following [6], we consider a combinatorial model of  $\Bbb R^4$
 as a certain 4-dimensional complex $C(4)$. Let $K(4)$ be a dual complex of
 $C(4)$. The complex $K(4)$ is a 4-dimensional complex of cochains with
 $su(2)$-valued coefficients. We define the discrete analogs of the connection
 1-form $A$ and the curvature 2-form $F$ as follows cochains
 $$
 A=\sum_k\sum_{i=1}^4A_k^ie_i^k, \qquad  F=\sum_k\sum_{i<j}\sum_{j=2}^4 F_k^{ij}\varepsilon_{ij}^k,\tag 9
 $$
 where \ $A_k^i, \ F_k^{ij}\in su(2)$, \ $e_i^k,  \ \varepsilon_{ij}^k$ \ are 1-,
 2-dimensional
 basis elements of $K(4)$ and $k=(k_1,k_2,k_3,k_4)$, $k_i\in\Bbb Z$.
 We use the geometrical construction proposed by A.~A.~Dezin in
 [1] to define discrete analogs of the differential, the exterior
 multiplication and the Hodge star operator.

 Let us introduce for convenient  the shifts operator $\tau_i$ and $\sigma_i$ as $$\tau_ik=(k_1,...\tau
 k_i,...k_4), \qquad
 \sigma_ik=(k_1,...\sigma k_i,...k_4),$$ where $\tau k_i=k_i+1$ and $\sigma k_i=k_i-1$,\ $k_i\in\Bbb Z$.
  Similarly, we denote by
 $\tau_{ij}$ ($\sigma_{ij}$) the operator shifting to the right (to the left) two differ components of $k=(k_1,k_2,k_3,k_4)$.
 For example, $\tau_{12}k=(\tau k_1,\tau k_2,k_3,k_4),$ $\sigma_{14}k=(\sigma k_1,k_2,k_3,\sigma k_4).$

 If we use (2) and take the definitions of $d$ and $\wedge$ in discrete case [1,6],
 then we obtain
 $$
 F_k^{ij}=\Delta_{k_i}A_k^j-\Delta_{k_j}A_k^i+A_k^iA_{\tau_ik}^j-
 A_k^jA_{\tau_jk}^i,\tag 10
 $$
 where $\Delta_{k_i}A_k^j=A_{\tau_ik}^j-A_k^j$, \ $i,j=1,2,3,4$.
 The metric adjoint operation $\ast$ acts on the 2-dimensional
 basis elements of $K(4)$ as follows
$$\aligned \ast\varepsilon_{12}^k&=\varepsilon_{34}^{\tau_{12}k},
\qquad \ast\varepsilon_{13}^k=-\varepsilon_{24}^{\tau_{13}k},
\qquad \ast\varepsilon_{14}^k=\varepsilon_{23}^{\tau_{14}k},\\
 \ast\varepsilon_{23}^k&=\varepsilon_{14}^{\tau_{23}k},
\qquad \ast\varepsilon_{24}^k=-\varepsilon_{13}^{\tau_{24}k},
\qquad
\ast\varepsilon_{34}^k=\varepsilon_{12}^{\tau_{34}k}.\endaligned
$$ Then we obtain $$ \ast F=\sum_k
\big(F_{\sigma_{34}k}^{34}\varepsilon_{12}^k-
F_{\sigma_{24}k}^{24}\varepsilon_{13}^k +
F_{\sigma_{23}k}^{23}\varepsilon_{14}^k+ $$ $$
\quad\qquad\qquad+F_{\sigma_{14}k}^{14}\varepsilon_{23}^k-
F_{\sigma_{13}k}^{13}\varepsilon_{24}^k +
F_{\sigma_{12}k}^{12}\varepsilon_{34}^k\big). \tag11 $$ Comparing
the latter and (9) the discrete analog of the self-dual equation
(the first equation of (8))
 we can written as follows
  $$\aligned
F_k^{12}&=F_{\sigma_{34}k}^{34},\qquad
F_k^{13}=-F_{\sigma_{24}k}^{24}, \qquad
F_k^{14}=F_{\sigma_{23}k}^{23},\\ F_k^{23}&=F_{\sigma_{14}k}^{14},
\qquad F_k^{24}=-F_{\sigma_{13}k}^{13}, \qquad
F_k^{34}=F_{\sigma_{12}k}^{12}
\endaligned\tag12
$$ for all $k=(k_1,k_2,k_3,k_4),$ \ $k_i\in\Bbb Z.$  Using (10)
Equations (12) can be rewritten in the following difference form:
$$\aligned &\Delta_{k_1}A_k^2-\Delta_{k_2}A_k^1+A_k^1\cdot
A_{\tau_1k}^2 -A_k^2\cdot A_{\tau_2k}^1=\\
&=\Delta_{k_3}A_{\sigma_{34}k}^4-\Delta_{k_4}A_{\sigma_{34}k}^3+
A_{\sigma_{34}k}^3\cdot A_{\sigma_4k}^4 -A_{\sigma_{34}k}^4\cdot
A_{\sigma_3k}^3,
\endaligned
$$

\smallskip
$$\aligned &\Delta_{k_1}A_k^3-\Delta_{k_3}A_k^1+A_k^1\cdot
A_{\tau_1k}^3 -A_k^3\cdot A_{\tau_3k}^1=\\
&=-\Delta_{k_2}A_{\sigma_{24}k}^4+\Delta_{k_4}A_{\sigma_{24}k}^2-
A_{\sigma_{24}k}^2\cdot A_{\sigma_4k}^4 +A_{\sigma_{24}k}^4\cdot
A_{\sigma_2k}^2,
\endaligned
$$

\smallskip
$$\aligned &\Delta_{k_1}A_k^4-\Delta_{k_4}A_k^1+A_k^1\cdot
A_{\tau_1k}^4 -A_k^4\cdot A_{\tau_4k}^1=\\
&=\Delta_{k_2}A_{\sigma_{23}k}^3-\Delta_{k_3}A_{\sigma_{23}k}^2+
A_{\sigma_{23}k}^2\cdot A_{\sigma_3k}^3 -A_{\sigma_{23}k}^3\cdot
A_{\sigma_2k}^2,
\endaligned
$$

\smallskip
$$\aligned &\Delta_{k_2}A_k^3-\Delta_{k_3}A_k^2+A_k^2\cdot
A_{\tau_2k}^3 -A_k^3\cdot A_{\tau_3k}^2=\\
&=\Delta_{k_1}A_{\sigma_{14}k}^4-\Delta_{k_4}A_{\sigma_{14}k}^1+
A_{\sigma_{14}k}^1\cdot A_{\sigma_4k}^4 -A_{\sigma_{14}k}^4\cdot
A_{\sigma_1k}^1,
\endaligned
$$

\smallskip
$$\aligned &\Delta_{k_2}A_k^4-\Delta_{k_4}A_k^2+A_k^2\cdot
A_{\tau_2k}^4 -A_k^4\cdot A_{\tau_4k}^2=\\
&=-\Delta_{k_1}A_{\sigma_{13}k}^3+\Delta_{k_3}A_{\sigma_{13}k}^1-
A_{\sigma_{13}k}^1\cdot A_{\sigma_3k}^3 +A_{\sigma_{13}k}^3\cdot
A_{\sigma_1k}^1,
\endaligned
$$

\smallskip
$$\aligned &\Delta_{k_3}A_k^4-\Delta_{k_4}A_k^3+A_k^3\cdot
A_{\tau_3k}^4 -A_k^4\cdot A_{\tau_4k}^3=\\
&=\Delta_{k_1}A_{\sigma_{12}k}^2-\Delta_{k_2}A_{\sigma_{12}k}^1+
A_{\sigma_{12}k}^1\cdot A_{\sigma_2k}^2 -A_{\sigma_{12}k}^2\cdot
A_{\sigma_1k}^1.
\endaligned
$$ In the same way we obtain the difference anti-self-dual
equation. From Equations (12) we obtain at once
 $$ F_k^{jr}=F_{\sigma
k}^{jr} \tag13 $$ for all  $j<r, \ r=2,3,4$, where $\sigma
k=(\sigma k_1,\sigma k_2,\sigma k_3,\sigma k_4)$.

Note that
Equations (13) also are satisfied in the case of the difference
anti-self-dual equations.
\bigskip
\noindent {\bf Proposition~1.}  {\it Let $F$ be a solution of the
discrete self-dual or anti-self dual equations. Then we have} $$
\ast\ast F= F.\tag14 $$
\smallskip
%\bigskip
\noindent {\bf Proof.} From (11) we have $$\aligned \ast\ast
F=&\sum_k\big(F_{\sigma_{34}k}^{34}\ast\varepsilon_{12}^k-
F_{\sigma_{24}k}^{24}\ast\varepsilon_{13}^k +
F_{\sigma_{23}k}^{23}\ast\varepsilon_{14}^k+\\&+
F_{\sigma_{14}k}^{14}\ast\varepsilon_{23}^k-
F_{\sigma_{13}k}^{13}\ast\varepsilon_{24}^k +
F_{\sigma_{12}k}^{12}\ast\varepsilon_{34}^k\big)=
\endaligned
$$ $$\aligned \qquad=&\sum_k
\big(F_{\sigma_{34}k}^{34}\varepsilon_{34}^{\tau_{12}k}+
F_{\sigma_{24}k}^{24}\varepsilon_{24}^{\tau_{13}k} +
F_{\sigma_{23}k}^{23}\varepsilon_{23}^{\tau_{14}k}+\\&+
F_{\sigma_{14}k}^{14}\varepsilon_{14}^{\tau_{23}k}+
F_{\sigma_{13}k}^{13}\varepsilon_{13}^{\tau_{24}k} +
F_{\sigma_{12}k}^{12}\varepsilon_{12}^{\tau_{34}k}\big)=\\=&
\sum\limits_k\sum\limits_{i<j}\sum\limits_{j=2}^4 F_{\sigma
k}^{ij}\varepsilon_{ij}^k.
\endaligned
$$ Comparing the latter and (13) we obtain (14).

 \hfill$\square$
\bigskip

 It should be noted that in the case of
continual Yang-Mills theory for $\Bbb R^4$ with the usual
Euclidean metric Equation (14) is satisfied automatically for an
arbitrary 2-form. But in the formalism we use the operation
$(\ast)^2$ is equivalent to a shift.

 The difference analog of Equations
(13) is given by $$\aligned
&\Delta_{k_j}A_k^r-\Delta_{k_r}A_k^j+A_k^j\cdot A_{\tau_jk}^r
-A_k^r\cdot A_{\tau_rk}^j=\\&=\Delta_{k_j}A_{\sigma
k}^r-\Delta_{k_r}A_{\sigma k}^j+ A_{\sigma k}^j\cdot
A_{\sigma\tau_jk}^r -A_{\sigma k}^r\cdot
A_{\sigma\tau_rk}^j,\endaligned$$ where $\sigma\tau_jk=(\sigma
k_1...k_j...\sigma k_4)$.

\smallskip
\bigskip
\noindent {\bf 3.~The discrete model in Minkowski space}
\bigskip

Let a base space of the bundle $P$ be Minkowski space, i.~e. $\Bbb R^4$
with the metric $g_{\mu\nu}=diag(-+++)$. In Minkowski space we write Equations (8)
as
$$
\ast F=\mp iF,\tag 15
$$
where $i^2=-1$. Recall that  $F$ is $\frak g$-valued, so therefore
is $\ast F$. Then we must have $i\frak g=\frak g$
in obvious notation. However, this latter condition is not satisfied
for the Lie algebras of any compact Lie groups $G$. To study Equations (15)
we must choose non-compact $G$ such as $SL(n,\Bbb C)$ or $GL(n,\Bbb C)$ say.
This is a serious restriction since in physics the gauge groups chosen
are usually compact [5]. Let the gauge group be $G=SL(2,\Bbb C)$.

We suppose that a combinatorial model of Minkowski space has the
same structure as $C(4)$. A gauge-invariant discrete model of the
Yang-Mills equations in Minkowski space is given in [7]. Now the
dual complex $K(4)$ is a complex of $sl(2,\Bbb C)$-valued cochains
(forms). Because discrete analogs of the differential and the
exterior multiplication are not depended on a metric then they
have the same form as in the case of Euclidean space. For more
details on this point see [7]. However, to define a discrete
analog of the $\ast$ operation we must take into accounts the
Lorentz metric structure on $K(4)$. We denote by $\bar x_{\kappa},
\ \bar e_{\kappa}$,  $\kappa\in\Bbb Z$ the basis elements of  the
1-dimensional complex $K$ which are corresponded to the time
coordinate of Minkowski space. It is convenient to write the basis
elements of $K(4)=K\otimes K\otimes K\otimes K$ in the form
$\bar\mu^{\kappa}\otimes s^k$, where  $\bar\mu^{\kappa}$ is either
$\bar x^{\kappa}$ or $\bar e^{\kappa}$ and $s^k$ is a basis
element of   $K(3)=K\otimes K\otimes K$,  $k=(k_1,k_2,k_3)$,
 \ $\kappa, k_j\in\Bbb Z$. Then we define the $\ast$ operation on
$K(4)$ as follows $$ \bar\mu^\kappa\otimes
s^k\cup\ast(\bar\mu^\kappa\otimes s^k)= Q(\mu)\bar e^\kappa\otimes
e^{k_1}\otimes e^{k_2}\otimes e^{k_3},\tag16 $$ where $Q(\mu)$ is
equal to $+1$ if $\bar\mu^\kappa=\bar x^\kappa$ and to $-1$ if
$\bar\mu^\kappa=\bar e^\kappa$. To arbitrary forms the $\ast$
operation is extended linearly. Using (16) we obtain
 $$ \ast F=\sum_k
\big(F_{\sigma_{34}k}^{34}\varepsilon_{12}^k-
F_{\sigma_{24}k}^{24}\varepsilon_{13}^k +
F_{\sigma_{23}k}^{23}\varepsilon_{14}^k- $$ $$
\quad\qquad\qquad-F_{\sigma_{14}k}^{14}\varepsilon_{23}^k+
F_{\sigma_{13}k}^{13}\varepsilon_{24}^k -
F_{\sigma_{12}k}^{12}\varepsilon_{34}^k\big),
 \tag17 $$
where
$F_k^{ij}\in sl(2,\Bbb C)$. Combining (17) with (9) the discrete
self-dual equation $\ast F=iF$ can be written as follows
$$\aligned
&F_{\sigma_{34}k}^{34}=iF_k^{12}, \qquad -F_{\sigma_{24}k}^{24}=iF_k^{13},
\qquad F_{\sigma_{23}k}^{23}=iF_k^{14},\\
-&F_{\sigma_{14}k}^{14}=iF_k^{23}, \qquad F_{\sigma_{13}k}^{13}=iF_k^{24},
\qquad -F_{\sigma_{12}k}^{12}=iF_k^{34}
\endaligned\tag18 $$
 for all $k=(k_1,k_2,k_3,k_4),
\ k_r\in\Bbb Z, \ r=1,2,3,4$. From the latter we obtain
 $$
F_{\sigma k}^{34}=iF_{\sigma_{12}k}^{12}=-i^2F_k^{34}=F_k^{34}, \qquad
F_{\sigma k}^{24}=-iF_{\sigma_{13}k}^{13}=-i^2F_k^{24}=F_k^{24} $$
and
similarly for any other components $F_k^{jr}, \ j<r$.
So we have Relations (13).
Thus a solution of the discrete self-dual equations (18) satisfies Equations (13)
as in the Euclidean case.

We can also rewrite (18) in the difference form

$$\aligned
\Delta_{k_3}A_{\sigma_{34}k}^4&-\Delta_{k_4}A_{\sigma_{34}k}^3+
A_{\sigma_{34}k}^3\cdot A_{\sigma_4k}^4
-A_{\sigma_{34}k}^4\cdot A_{\sigma_3k}^3=\\
&=i(\Delta_{k_1}A_k^2-\Delta_{k_2}A_k^1+A_k^1\cdot A_{\tau_1k}^2
-A_k^2\cdot A_{\tau_2k}^1),
\endaligned
$$

\smallskip
$$\aligned
 -\Delta_{k_2}A_{\sigma_{24}k}^4&+\Delta_{k_4}A_{\sigma_{24}k}^2-
A_{\sigma_{24}k}^2\cdot A_{\sigma_4k}^4
+A_{\sigma_{24}k}^4\cdot A_{\sigma_2k}^2=\\
&=i(\Delta_{k_1}A_k^3-\Delta_{k_3}A_k^1+A_k^1\cdot A_{\tau_1k}^3
-A_k^3\cdot A_{\tau_3k}^1),
\endaligned
$$

\smallskip
$$\aligned
\Delta_{k_2}A_{\sigma_{23}k}^3&-\Delta_{k_3}A_{\sigma_{23}k}^2+
A_{\sigma_{23}k}^2\cdot A_{\sigma_3k}^3
-A_{\sigma_{23}k}^3\cdot A_{\sigma_2k}^2=\\
&=i(\Delta_{k_1}A_k^4-\Delta_{k_4}A_k^1+A_k^1\cdot A_{\tau_1k}^4
-A_k^4\cdot A_{\tau_4k}^1),
\endaligned
$$

\smallskip
$$\aligned
-\Delta_{k_1}A_{\sigma_{14}k}^4&+\Delta_{k_4}A_{\sigma_{14}k}^1-
A_{\sigma_{14}k}^1\cdot A_{\sigma_4k}^4
+A_{\sigma_{14}k}^4\cdot A_{\sigma_1k}^1=\\
&=i(\Delta_{k_2}A_k^3-\Delta_{k_3}A_k^2+A_k^2\cdot A_{\tau_2k}^3
-A_k^3\cdot A_{\tau_3k}^2),
\endaligned
$$

\smallskip
$$\aligned
 \Delta_{k_1}A_{\sigma_{13}k}^3&-\Delta_{k_3}A_{\sigma_{13}k}^1+
A_{\sigma_{13}k}^1\cdot A_{\sigma_3k}^3
-A_{\sigma_{13}k}^3\cdot A_{\sigma_1k}^1=\\
&=i(\Delta_{k_2}A_k^4-\Delta_{k_4}A_k^2+A_k^2\cdot A_{\tau_2k}^4
-A_k^4\cdot A_{\tau_4k}^2),
\endaligned
$$

\smallskip
$$\aligned
-\Delta_{k_1}A_{\sigma_{12}k}^2&+\Delta_{k_2}A_{\sigma_{12}k}^1-
A_{\sigma_{12}k}^1\cdot A_{\sigma_2k}^2
+A_{\sigma_{12}k}^2\cdot A_{\sigma_1k}^1=\\
&=i(\Delta_{k_3}A_k^4-\Delta_{k_4}A_k^3+A_k^3\cdot A_{\tau_3k}^4
-A_k^4\cdot A_{\tau_4k}^3).
\endaligned
$$

\smallskip
In similar manner we obtain the difference anti-self-dual
equations. Obviously an anti-self-dual solution satisfies
Equations (13).

\bigskip
\noindent {\bf Proposition~2.}  {\it Let for any $sl(2,\Bbb C)$-valued
2-form $F$ Conditions (13)
are satisfied. Then we have}
 $$ \ast\ast F=- F. $$

\bigskip
\noindent {\bf Proof.} If components of any discrete 2-form $F$
satisfy (13), then $F$ is a solution of the discrete self-dual or
anti-self-dual equations. Hence $$ \ast\ast F=\ast(\mp iF)=\mp
i\ast F=(\mp i)^2F=-F. $$

\hfill$\square$
\bigskip
\noindent {\bf Remark.} {\it In the continual case
the self-dual and anti-self-dual equations are written in the form (15)
because we have $\ast\ast F=-F$ for an arbitrary 2-form $F$ in
Minkowski space. In the
discrete model case it is easy to check that in $K(4)$ we
have
$$\ast\ast
F=-\sum\limits_k\sum\limits_{i<j}\sum\limits_{j=2}^4 F_{\sigma
k}^{ij}\varepsilon_{ij}^k.$$
 Thus  Equations (15) are satisfied only under  Conditions (13).}

\bigskip
\noindent {\bf Theorem.}  {\it If exist some   $N=(N_1,N_2,N_3,N_4), \  N_r\in \Bbb Z$
such that
$$
F_k^{ij}=0 \quad for \  any \quad  |k|\geq|N|,\tag19
$$
then Equations (15) (or (8)) have only the trivial solution $F=0$.}

\bigskip
\noindent
 {\bf Proof.} Since for any solution of Equations (15) (or (8)) we have
 Relations (13) then the assertion is obvious.

\hfill$\square$

Let $g$ be a discrete $0$-form
$$
g=\sum_k g_kx^k,
$$
where $x^k$ is the 0-dimensional basis element of $K(4)$ and $g_k\in SU(2)$
(or $g_k\in Sl(2,\Bbb C)$). The boundary condition (19) in terms of the connection components
can be represented as:
 there is some discrete 0-form $g$  such that
$$
A_k^j=-(\Delta_{k_j}g_k)g_k^{-1} \quad \text{for any} \quad   |k|\geq |N|.
$$
It follows from Theorem 3 [6].


\Refs \ref \no1 \by A. A. Dezin \book Multidimensional  analysis
and discrete models  \yr 1990 \publ Nauka, Moscow
\endref

\ref \no2 \by A. A. Dezin \paper Models connected with Yang-Mills
equations \jour Differentsial'nye Uravneniya \yr 1993 \vol 29
\issue 5 \pages 846--851  \transl\nofrills  English transl.  in
\jour Differential Equations \vol 29 \yr 1993 \endref
 \ref \no3 \by B. A. Dubrovin, S. P. Novikov, A. T. Fomenko
 \book Modern geometry \yr 1979 \publ Nauka, Moscow
\endref
 \ref \no4
\by D. S. Freed, K. K. Uhlenbeck \book Instantons and
four-manifolds \yr 1984 \publ Springer-Verlag New York Ins.
\endref
\ref \no5 \by C. Nash, S. Sen \book Topology and geometry for
physicists \publ Academic press London \yr 1989 \endref

\ref \no6 \by V. N. Sushch \paper Gauge-invariant discrete models
of Yang-Mills equations \jour Mat. Zametki \yr 1997 \vol 61 \issue
5 \pages 742--754 \transl\nofrills English transl.  in \jour
Mathematical Notes \vol 29 \yr 1997 \pages 621--631
\endref
\ref \no7 \by V.~N.~Sushch \paper On some discretization of
Yang-Mills equations in Minkowski space
 \jour Nonlinear boundary value problems\yr 2003  \issue
13 \pages 197--208
\endref
\endRefs
\enddocument